\documentclass[twocolumn]{aastex62}
\usepackage{xfrac}
\usepackage{amsmath}

\submitjournal{ApJ}

\shorttitle{Origin of Dark Gaps}
\shortauthors{Gnedin}

\def\dim#1{{\rm #1}}

\begin{document}
\title{Cosmic Reionization on Computers: Physical Origin of Long Dark Gaps in Quasar Absorption Spectra}

\correspondingauthor{Nickolay Y.\ Gnedin}
\email{gnedin@fnal.gov}

\author{Nickolay Y.\ Gnedin}
\affiliation{Fermi National Accelerator Laboratory;
Batavia, IL 60510, USA}
\affiliation{Kavli Institute for Cosmological Physics;
The University of Chicago;
Chicago, IL 60637 USA}
\affiliation{Department of Astronomy \& Astrophysics; 
The University of Chicago; 
Chicago, IL 60637 USA}

\begin{abstract}
I explore the properties of "dark gaps" - regions in quasar absorption spectra without significant transmission - with several simulations from the Cosmic Reionization On Computers (CROC) project. CROC simulations in largest available boxes (120 cMpc) come close to matching both the distribution of mean opacities and the frequency of dark gaps, but alas not in the same model: the run that matches the mean opacities fails to contain enough dark gaps and vice versa.:( Never-the-less, the run that matches the dark gap distributions serves as a counter-example to claims in the literature that the dark gap statistics requires a late end to reionization - in that run reionization ends at $z=6.7$ (likely too early).

While multiple factors contribute to the frequency of large dark gaps in the simulations, the primary factor that controls the overall shape of the dark gap distribution is the ionization level in voids - the lowest density regions produce the highest transmission spikes that terminate long gaps. As the result, the dark gap distribution correlates strongly with the fraction of the spectrum above the gap detection threshold, the observed distribution is matched by the simulation in which this fraction is 2\%. Hence, the gap distribution by itself does not constrain the timing of reionization.
\end{abstract}


\section{Introduction}
\label{sec:intro}

Absorption spectra of high-redshift quasars offer the most detailed, and thus the most constraining, probe of cosmic reionization currently available. Matching these observations is a challenge to modern simulations of  reionization. One of the steep challenges is the distribution of mean opacities in short skewers \citep{Becker2015,Bosman2018,Eilers2018,Yang2020}, which two of the recent most advanced large-scale reionization projects, "Cosmic Reionization On Computers" \citep[CROC,][]{gnedin14,gnedin_etal17} and THESAN \citep{Kannan2021,Garaldi2021} pass only marginally at best. And while matching the mean of an observed distribution is often the easiest task for a model, achieving a match for the tails of the distribution is usually way harder.

Such a challenge has been recently highlighted by \citet{Zhu2021}, who assembled a large sample of "dark gaps" (continuous spectral regions of low enough flux) in the quasar absorption spectra at redshifts as high as $z=6$. The existence of large gaps in excess of 200 comoving megaparsecs (hereafter cMpc) has been interpreted as evidence for late "completion" of reionization, perhaps as late as $z=5.3$. 

The concept of "the end of reionization" is inherently ambiguous and definition-dependent, and thus not particularly useful. For example, around 2\% of gas remains neutral after reionization as Damped Lyman-$\alpha$ (DLA) and sub-DLA systems \citep{Sanchez-Ram2016,Berg2019} and that fraction falls to about 0.5-1\% by $z=0$. Thus, if one defines "the end of reionization" as the moment when the mean (mass-weighted) neutral fraction falls below 2\%, the reionization "ends" around $z=3$; if the threshold of 1\% is chosen, it ends sometime between $z=1$ and $z=0$.

\begin{figure*}[t]
\includegraphics[width=0.5\hsize]{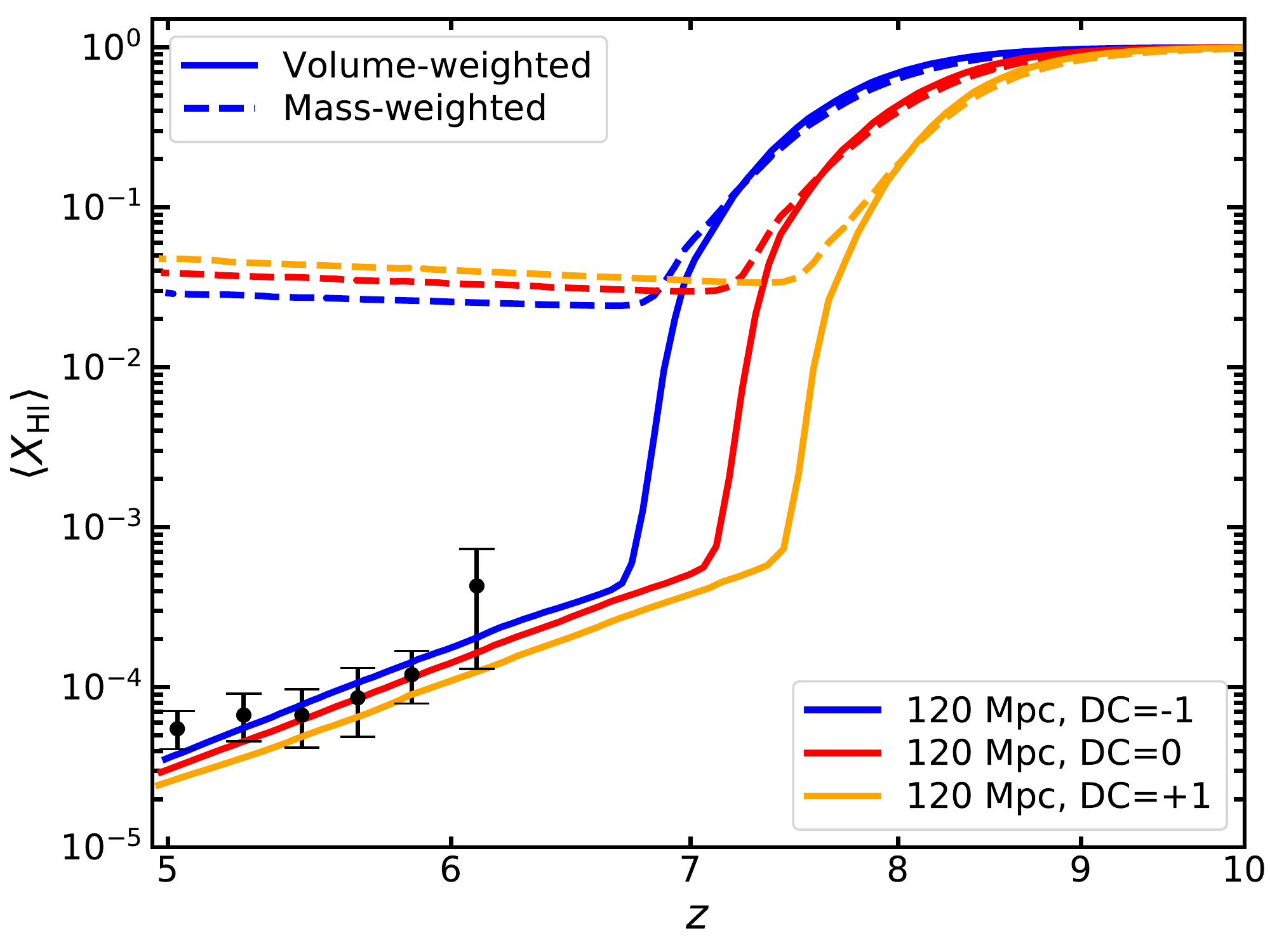}%
\includegraphics[width=0.5\hsize]{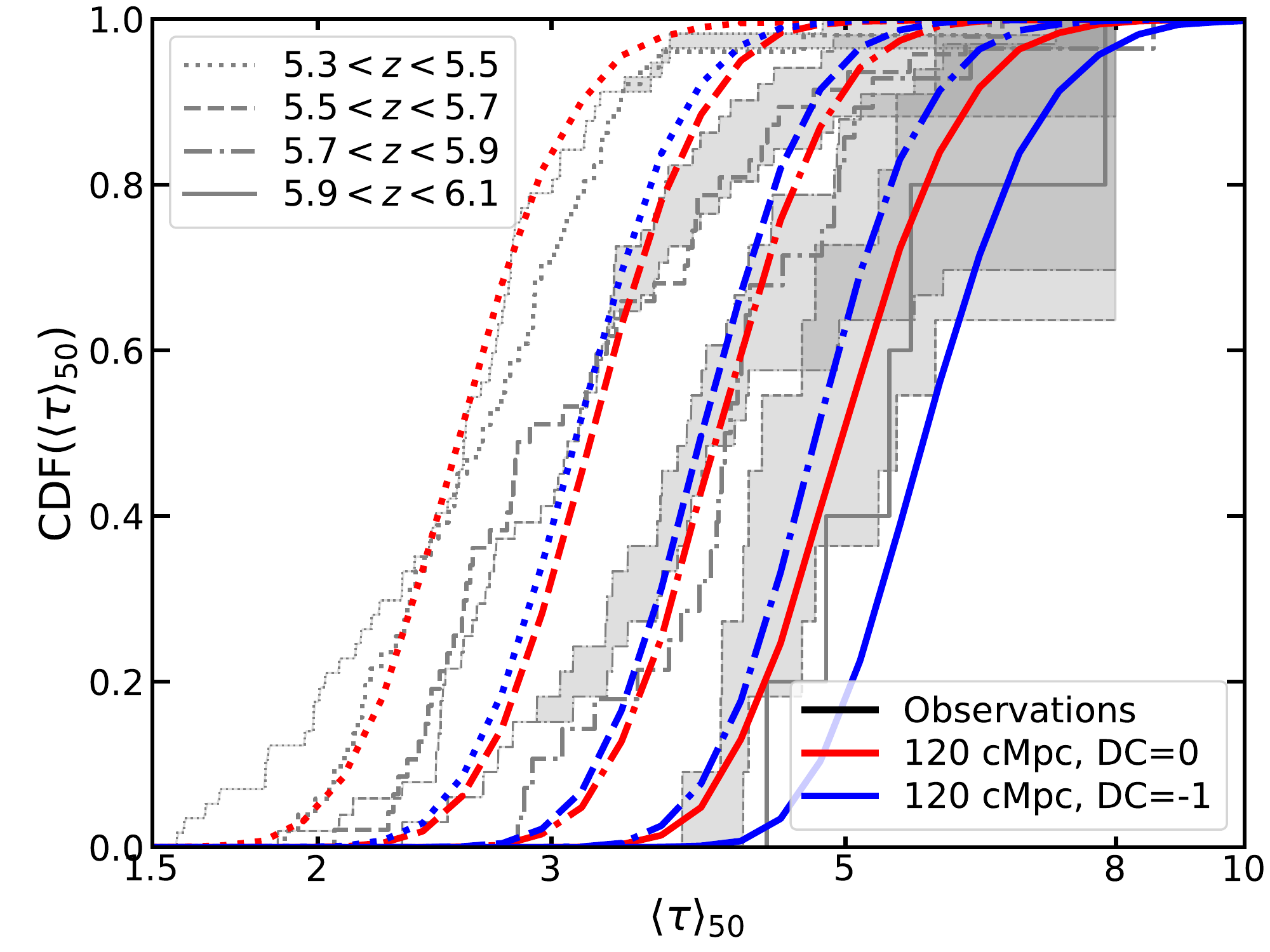}
\caption{(Left) Evolution of the volume-weighted (solid lines) and mass-weighted (dashed lines) fractions of neutral hydrogen for 3 different realizations of the $120\,\dim{cMpc}$ simulation volume used in this paper. The data points are from \citet{Fan2006} and were used as the calibration data for the simulations. (Right) Distributions of mean opacities in $50h^{-1}\dim{cMpc}$ skewers for the two realizations DC=0 and DC=-1. Gray lines and bands show the observational data from \citet{Becker2015} and \citet{Bosman2018}. The mean density model DC=0 (red lines) marginally matches the distribution of opacities (at least at $z>5.5$), while the DC=-1 model (blue lines) overpredicts opacities substantially. The DC=+1 model (not shown) underpredicts the opacity by about the same amount.} 
\label{fig:xhz}
\end{figure*}

In order to be quantitative, it is useful to come up with a physically-motivated definition for the "end of reionization". One such definitions (which I adopt here) is based on the generic evolution of the mean free path for ionizing photons: during reionization the mean free path is limited by the average size of an ionized bubble, and after reionization it is limited by the abundance of Lyman limit systems \citep[see, e.g., Fig.\ 8 in][]{Fan2006}. Hence, one can say that reionization is complete when the photon mean free path becomes limited by the Lyman limit systems. Another advantage of this definition (in addition to it being physically motivated rather than being based on an arbitrary threshold) is that it relates directly to the also physically motivated definition for the "redshift of overlap of ionized bubbles", the moment when the topology changes from  ionized-bubbles-in-the-neutral-universe to neutral-bubbles-in-the-ionized-universe, which is defined as the moment when the mean free path increases at the fastest rate \citep{Gnedin2000,Gnedin2004}. With these definitions reionization ends soon after the overlap of ionized bubbles.

Arguably an even more useful comparison is to do the full forward modeling of a simulation and compare the actual observed distribution of dark gaps with the simulated one. This is the main approach I adopt in this paper.

\section{Simulations and Data}

The primary simulation set immediately available to me is CROC simulations. In these paper I use 3 largest CROC  runs in simulation volumes with size of $80h^{-1}\dim{cMpc}\approx120\,\dim{cMpc}$ and with the mass resolution of $2048^3$ \citep{gnedin14,gnedin_etal17}. For reference, these simulations are similar in most respects to the largest THESAN box \citep{Kannan2021,Garaldi2021} and represent the largest currently achievable simulations of reionization (i.e.\ requiring in excess of 50 million core-hours per run).

The 3 simulations differ in the value of the mean density in the box, the so-called "DC mode" \citep{Pen1997,Sirko2005,Gnedin2011}. The mean density in the box is not zero, since a simulation box is finite and hence the density still fluctuates at the scale of the box. The first run (I call it "DC=0" hereafter) is one of the several true random realizations of a $80h^{-1}\dim{cMpc}$ simulation box, and is selected because by chance it has the mean density close to the true cosmic mean (the value of the DC mode, the fluctuation in the cosmic mean density at $z=0$, of 0.072). The other two boxes start from the same realization of the initial conditions but have the DC mode set to +1 and -1 manually. The rms density fluctuation in cubes of $80h^{-1}\dim{cMpc}$ at $z=0$ is 0.245, so these two runs represent $\pm4\sigma$ fluctuations, while the "DC=0" run is a $+0.3\sigma$ fluctuation.

Because the DC=$\pm$1 runs are such rare density fluctuations, they have significantly different reionization histories and can be interpreted as different reionization models. Figure \ref{fig:xhz} shows their reionization histories as represented by the evolution of the neutral hydrogen fraction. The behavior of the volume-weighted neutral fraction is characteristic - it decreases at an accelerating rate, reflecting the accelerated expansion of ionized bubbles, until the photon mean free path becomes limited by the Lyman limit systems, after which the mean neutral fraction decreases in a steady, power-law-like manner, reflecting the power-law like evolution of the Lyman limit systems. In the 3 runs this happens at $z=6.7$ (DC=-1), $z=7.1$ (DC=0), and $z=7.4$ (DC=+1) respectively.

The right panel shows comparison of two of the CROC runs (DC=0 and DC=-1) with the measured mean opacities in  $50h^{-1}\dim{cMpc}$ skewers from \citet{Becker2015} and \citet{Bosman2018}. The DC=0 model offers a marginally acceptable fit to the data (except perhaps at $z=5.6$), while the DC=-1 model overpredicts the opacities by about $\Delta\tau=0.7$. The DC=+1 model (not shown) underpredicts the opacities by about the same amount. 

The purpose of this paper is to explore the properties of dark gaps in CROC simulations. In order to directly compare with the data of \citet{Zhu2021}, I mimic their dark gap definition as close as realistically plausible. At each simulation snapshot 1000 randomly oriented lines of sight are produced with length of $200h^{-1}\dim{cMpc}$. This is 2.5 times the box size, and is a safe length for a line of sight to use for a simulation box with periodic boundary conditions \citep{DallAglio2010}. The lines of sight are then binned to $1h^{-1}\dim{cMpc}$ resolution and a Gaussian noise is added. The added noise samples the actual distribution of signal-to-noise ratios of the observational data \citep[Table 1 of][]{Zhu2021}. As I show in the appendix, the choice of $1h^{-1}\dim{cMpc}$ binning makes the effect of the noise negligible.

\section{Results}

\subsection{CROC "as is"}

\begin{figure}[t]
\includegraphics[width=\hsize]{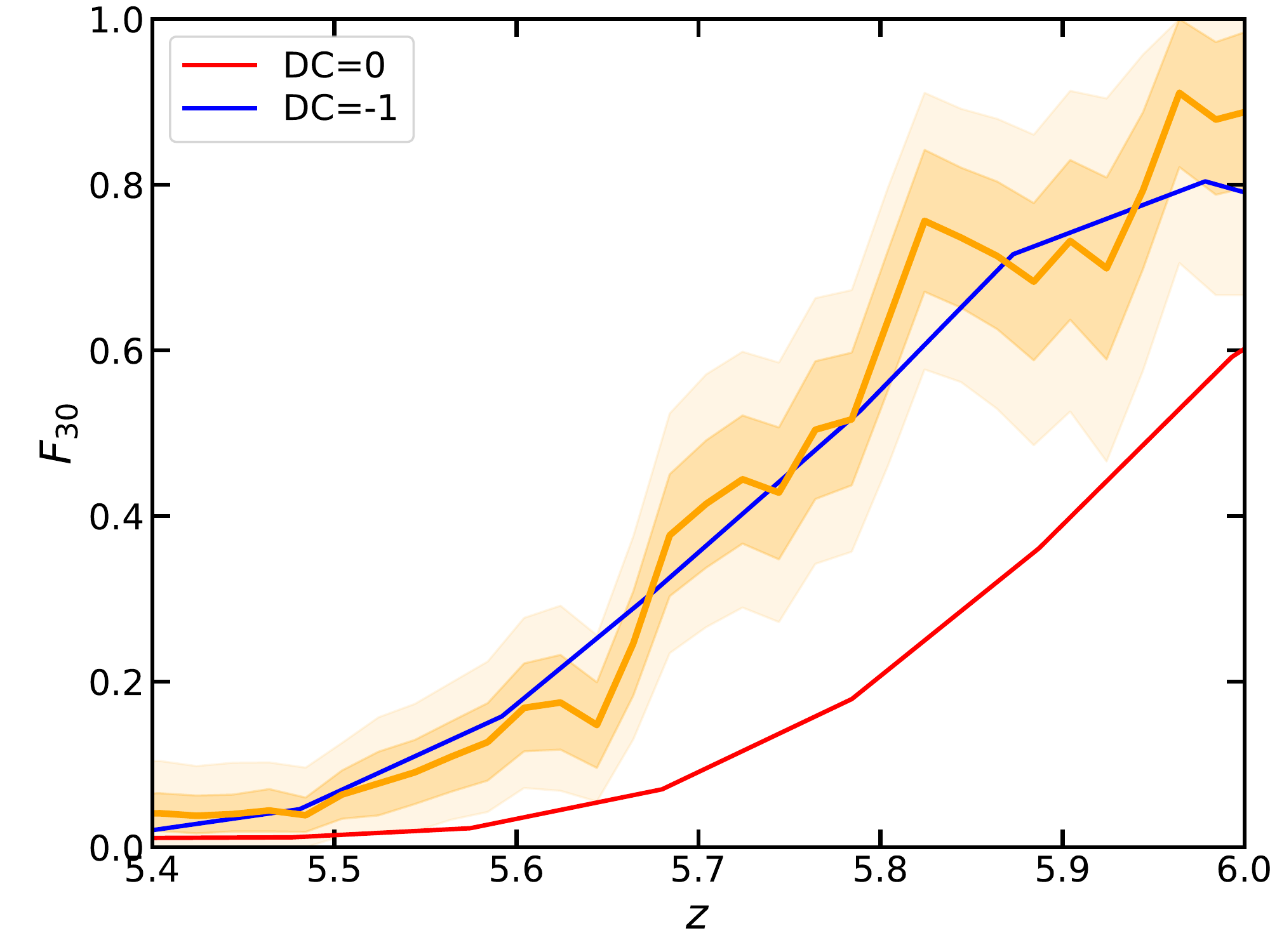}
\caption{Fraction of lines of sight exhibiting long ($L>30 h^{-1}\dim{cMpc}$) dark gaps as a function of redshift. The orange solid line with bands is the observational data from \citet{Zhu2021} with 68 and 95 percentile ranges. Other 2 lines are the same quantity estimated from the two CROC simulations as shown on the legend. The distribution of dark gaps is well sampled in the simulated data, so it is not necessary to bootstrap the synthetic spectra.}
\label{fig:f30}
\end{figure}

Figure \ref{fig:f30} shows the comparison of the 2 CROC runs (DC=0 and DC=-1) with Fig.\ 5 from \citet{Zhu2021}. For the primary observational diagnostic for the abundance of gaps \citet{Zhu2021} chose the fraction $F_{30}$ of sight lines that, at a given value of redshift $z$, intersect a dark gap with length in excess of $30 h^{-1}\dim{cMpc}$. This particular diagnostic is sub-optimal, because it is not easily reproducible in the simulations unless past light cones can be constructed - the largest gaps span substantial redshift ranges ($100 h^{-1}\dim{cMpc}$ is about $\Delta z=0.35$ at $z=6$). As a proxy for this diagnostic for a fixed simulation snapshot, I find all gaps that cross the specific distance from the origin of a synthetic line of sight (taken to be the half of the computational box size). Notice that it is not correct to just count the gaps of sufficient length in the synthetic spectra at a given snapshot; e.g.\ a line of sight A may have a gap of, say, $40 h^{-1}\dim{cMpc}$, at a distance of $30 h^{-1}\dim{cMpc}$ from the origin, while a line of sight B may have a similar gap at a distance of $90 h^{-1}\dim{cMpc}$ from the origin. If these two lines of sight were sitting on the same past light cone, the redshift ranges corresponding to the beginning and end of both gaps ($30-70$ and $90-130$) would not overlap, and hence only one of these gaps should be counted in the $F_{30}$ diagnostic.

The fiducial CROC run (DC=0) grossly underpredicts the observed number of gaps, but the later reionizing one (DC=-1) however shows a good agreement with the data. Note that neighboring values for the $F_{30}$ diagnostic are highly correlated over the redshift range of at least $0.12$, corresponding to the gap length of $30 h^{-1}\dim{cMpc}$, with some correlation persisting to $\Delta z\sim 0.4$. The early reionizing run (DC=+1) fails to match either the distribution of optical depths or the $F_{30}$ diagnostic.

\begin{figure*}[t]
\includegraphics[width=0.5\hsize]{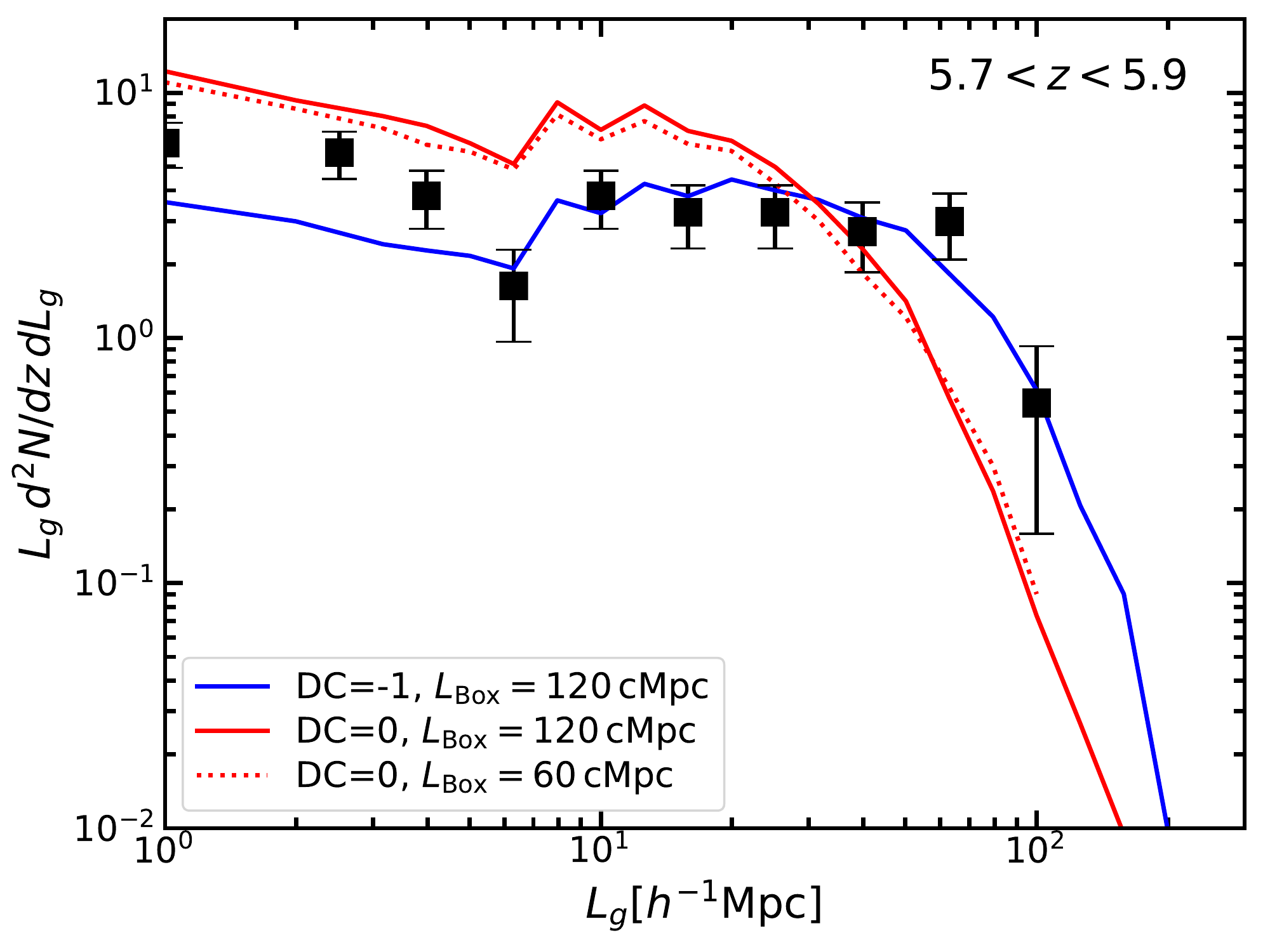}%
\includegraphics[width=0.5\hsize]{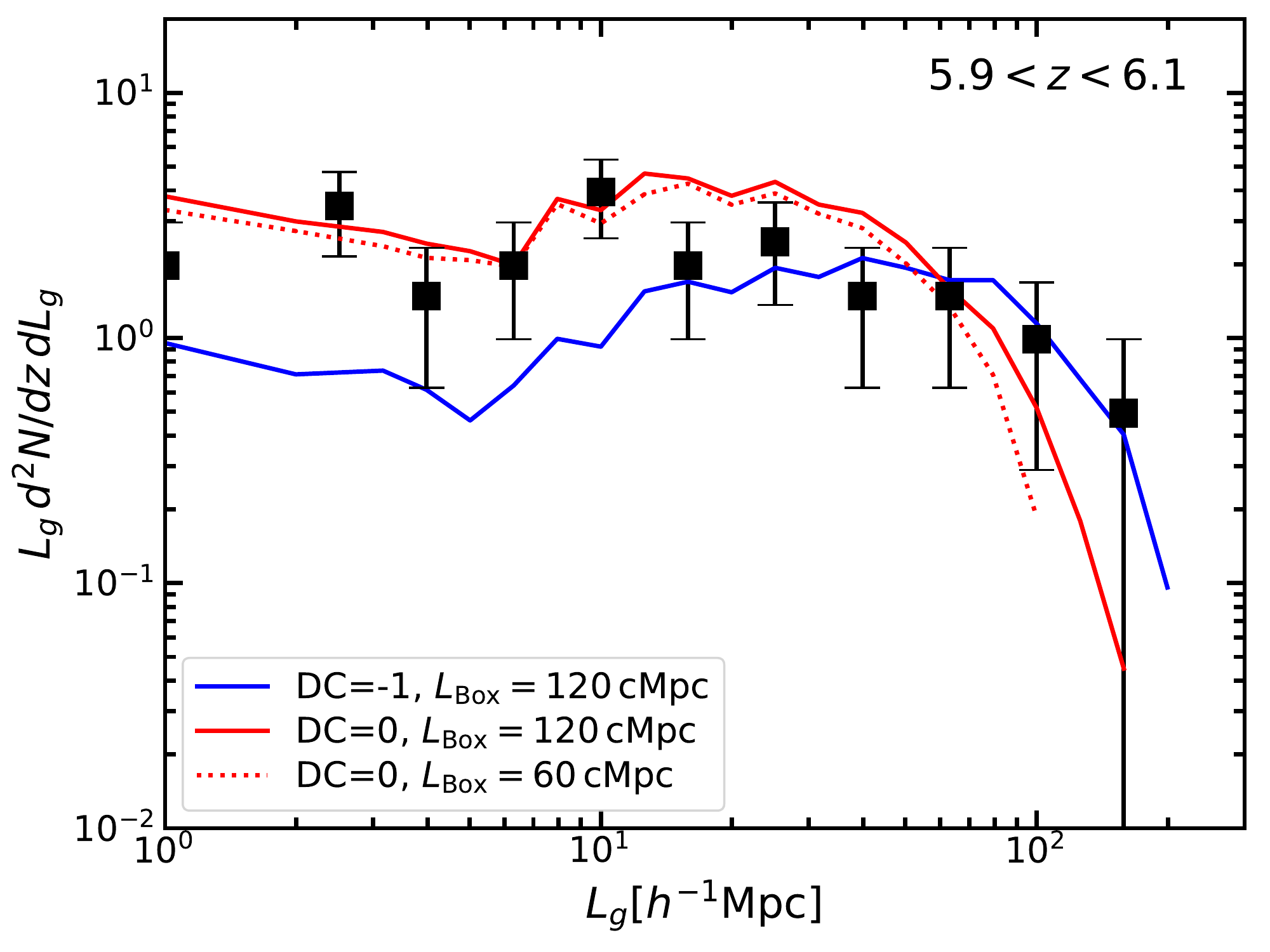}
\caption{Probability distribution function for gaps of different lengths per unit redshift for gaps whose central redshifts fall in the specified redshift bins ($[5.7,5.9]$ and $[5.9,6.1]$). Black squares are observational data from \citet{Zhu2021} with error bars computed by bootstrapping. Blue and red lines show the DC=-1 and DC=0 CROC simulations respectively, as well as the gap distribution from a set of 4 smaller boxes, as a test of numerical convergence with respect to the simulation box size.}
\label{fig:gaps}
\end{figure*}

For a more quantitative comparison, I show in Figure \ref{fig:gaps} the probability distribution function for gaps of different lengths whose central redshifts fall in the redshift bins $5.7<z<5.9$ and $5.9<z<6.1$. This quantity is more "simulation friendly" as it can be computed from a single snapshot without the need to generate light cones. Its another advantage is that different redshifts bins are independent. 

The DC=-1 run again provides a decent match to the data for sufficiently large gaps, although it fails to match the abundance of gaps smaller than about $10h^{-1}\dim{cMpc}$, for which the DC=0 run offers a better match. This is likely related to the level of agreement between these two runs and the observed distribution of optical depths shown in Fig.\ \ref{fig:xhz}, but with just two runs that cannot be easily repeated it is unclear how to investigate that relation further. 

Since the sizes of the simulation volumes I have to work with are comparable to the sizes of the largest observed gaps, the limited simulation volume may introduce a bias. For checking its effect, I also show in Fig.\ \ref{fig:gaps} the gap distribution from 4 random realizations of a $40h^{-1}\dim{cMpc}\approx60\,\dim{cMpc}$ box. These 4 realizations have random values of their DC modes but are selected so that the average DC mode for 4 of them is close to 0 and hence these 4 runs combined can be meaningfully compared with the DC=0 $80h^{-1}\dim{cMpc}$ run. The effect of the simulation box size is small - its main role is in restricting the maximum size of the synthetic dark gap to the length of the synthetic spectrum. This is fully consistent with the simulations presented in \citet{Zhu2021}. I also confirm (using a $40h^{-1}\dim{cMpc}$ run with 8 times higher mass and 2 times higher spatial resolution simulation) their conclusion that the spatial resolution of the simulation does not substantially affect the distribution of dark gaps - this is fully expected, as all simulations used here and in \citet{Zhu2021} have spatial resolution much higher than the smoothing scale of $1h^{-1}\dim{cMpc}$ used to detect gaps. I also checked that the choice of the pixel size in the synthetic spectra is unimportant as long as it is $10\dim{km/s}$ or below.

\subsection{Model for gaps}

\begin{figure}[t]
\includegraphics[width=\hsize]{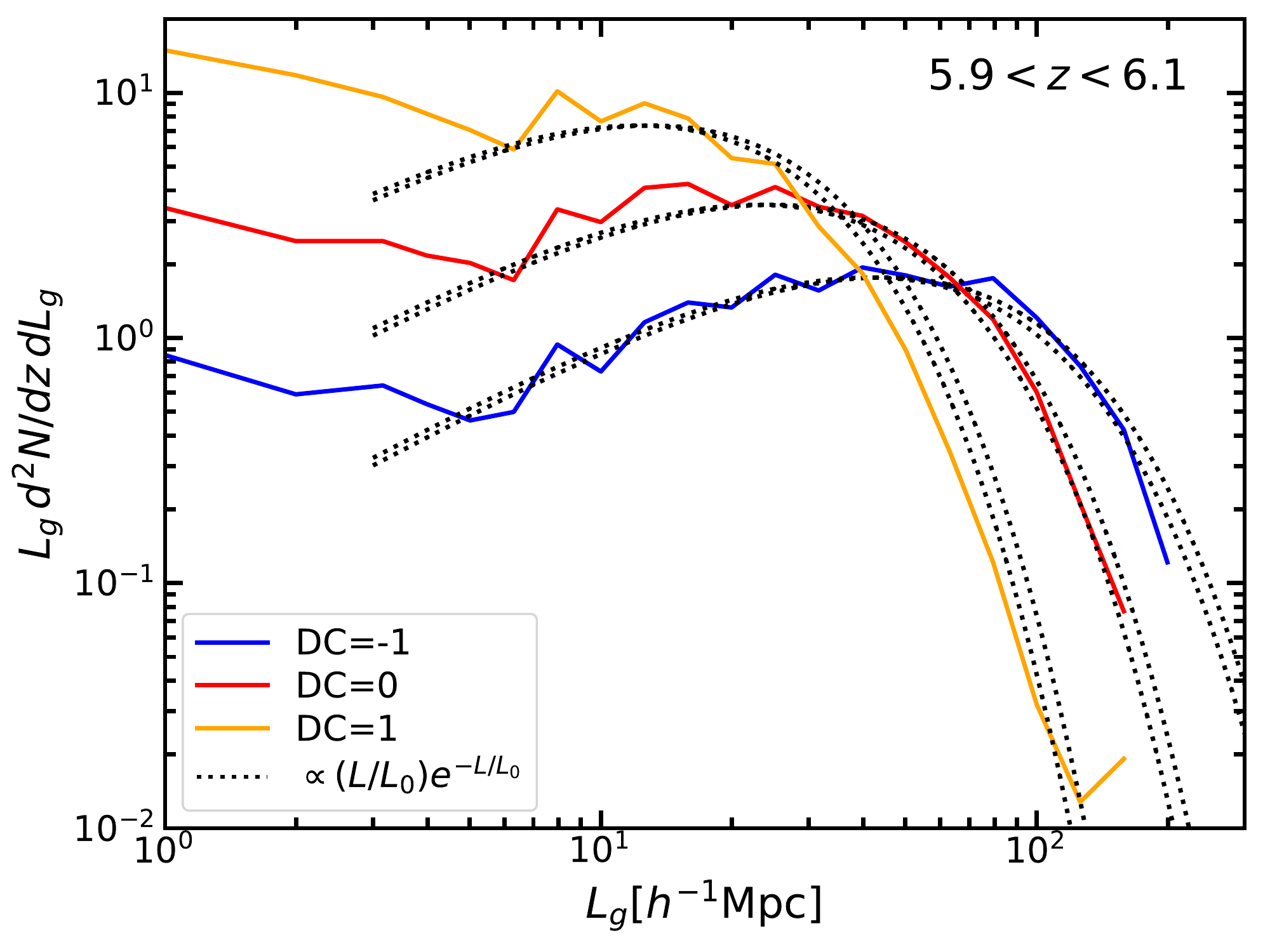}%
\caption{Probability distribution function for gaps of different lengths per unit redshift for the DC=-1 and DC=0 CROC simulations (as in Fig.\ \ref{fig:gaps} but without noise). For each simulation I show a distribution $L/L_0e^{-L/L_0}$ of line segments that can be placed in between randomly distributed hard spheres, with two different values of $L_0$ for each run that correspond to two commonly used variations of the void abundance model, as described in the text.}
\label{fig:gapsfit}
\end{figure}

So, what determines the sizes of the largest gaps? Figure \ref{fig:gapsfit} may give a hint: it shows that the gap distribution is well fitted by a distribution of sizes of line segments that are terminated by randomly distributed solid spheres,
\[
    L\frac{dN}{dL} \propto \frac{L}{L_0} e^{-L/L_0},
\]
where $L_0 = 1/(\pi R^2 n)$, $R$ is the sphere radius and $n$ is the number density of spheres. Since the transmission spikes (that terminate gaps) come primarily from the lowest density regions (i.e.\ voids), the line of sight contains a high enough spike (which terminates the dark gap) when it crosses a deep enough void. The fact that the gap distributions in simulations follow $L/L_0\exp(-L/L_0)$ very well implies that the effect of voids can be represented as a random distribution of hard spheres (voids are known to cluster little on $100h^{-1}\dim{Mpc}$ scales, \citep{Hamaus2014b,Chan2014,Hamaus2016}).

Most of voids originate from the downward fluctuations in the initial Gaussian field, and because the Gaussian field is symmetric with respect to the sign flip, one can use the known formulae for the number density of peaks of various height \citep{BBKS} to estimate the number density of voids that originate from the negative fluctuations of a given amplitude $\nu = -\delta/\sigma$ (with $\delta<0$). However, in order to compute which void can produce a high enough spike and which cannot, one would need to build a model for the ionization state inside a void to compute $R$. That is not impossible, since voids are known to have universal profiles \citep{Hamaus2014}, but is not a small undertaking either. A much easier, although a less enlightening approach, is to use the fraction of spectral pixels $f_{>}$ above the flux threshold chosen for the definition of a gap - for the DC=-1 run that value is 2.1\% and for the DC=0 and DC=1 runs this fraction is 5.1\% and 13.2\% respectively. 

Not all downward fluctuations produce voids, though - this is known as the "void-in-cloud" problem \citep{Sheth2004,Jennings2013}, as small voids can be squashed and incorporated into overdense structures. \citet{Sheth2004} give an expression for the fraction of downward density fluctuations that do result in voids (their Fig.\ 7),
\[
  f_{\rm void}(\nu) = \exp\left(-\frac{D}{1-D}\frac{D^2}{4\nu^2}-2\frac{D^4}{\nu^4}\right)
\]
with $D=\delta_v/(\delta_v+\delta_c)$, $\delta_c=1.69$ is the linear collapse threshold in the EdS universe (which is applicable to high redshift), and $\delta_v$ is the characteristic void underdensity. \citet{Sheth2004} used the value of $\delta_v=2.81$, but later works found that a lower values of $\delta_v\approx1$ matches numerical simulations better \citep{delv1,delv2}.

For the number density of peaks as a function of $\nu$ I take $n_{\rm peak}(\nu)$ from \citet{BBKS} with the LCDM power spectrum at $z=6$ smoothed with a Gaussian filter of FWHM of $1h^{-1}\dim{Mpc}$ (to approximately match the observational selection); integrating it with the fraction of downward fluctuations resulting in voids produces the number density of voids
\[
  n_{\rm void} = \int\limits_0^\infty n_{\rm peak} f_{\rm void} d\nu = 0.0144(0.0183) h^3\dim{Mpc}^{-3}
\]
for $\delta_v=2.81 (1.0)$ respectively. The same value for the number density of voids is obtained if I simply take all voids with $\nu \geq 1(0.6)$.

For a given fraction of spectral pixels above the threshold $f_{>}$, the central region of radius $R$ is ionized enough to produce the flux above threshold for $R=(3f_{>}/(4\pi n_{\rm void})^{1/3}$. The size of this region may be $\nu$-dependent, and in that case the value of $R$ is the average one. Given $R$, the characteristic gap length $L_0 = 1/(\pi R^2 n_{\rm void})$. For the 3 values of $f_{>}$ quoted above for the three CROC runs, the values for the characteristic gap length are $L_0\approx 44.6(41.2)h^{-1}\dim{cMpc}$, $24.7(22.8)h^{-1}\dim{cMpc}$, and $13.1(12.1)h^{-1}\dim{cMpc}$ for the DC=-1, DC=0, and DC=1 runs respectively (values in parenthesis are again for $\delta_v=1$). These are the values used for black dotted lines in Fig.\ \ref{fig:gapsfit}.

It is possible that the good agreement between the simple calculation above and the simulation results is just a coincidence, but then it is a triple one.

\begin{figure}[t]
\includegraphics[width=\hsize]{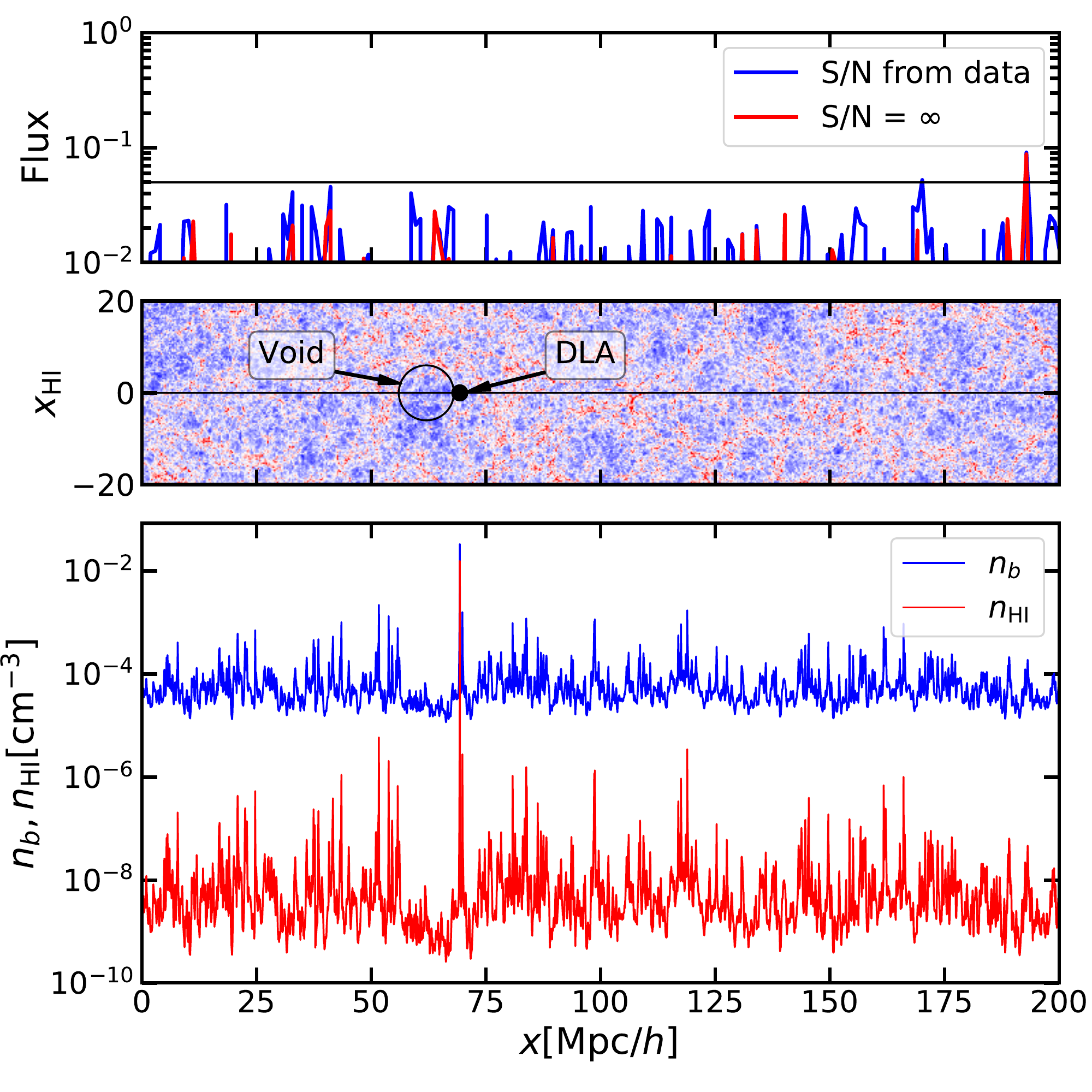}
\caption{Example of a synthetic line of sight containing a dark gap of length $192h^{-1}\dim{cMpc}$. The bottom panel shows the profiles of the total baryon number density (blue) and neutral hydrogen number density (red). This line of sight crosses a Damped Lyman-$\alpha$ system (DLA) at $x\approx70h^{-1}\dim{cMpc}$. The middle panel shows the slice through the computational volume at a random orientation perpendicular to the synthetic line of sight. The slice shows the gas density with the logarithmic stretch between $1+\delta=0.1$ (blue) to $1+\delta=10$ (red) and with higher/lower values capped to the values at the bounds (10/0.1 respectively). The line of sight mostly goes along the filaments, avoiding large voids (blue colors), except a part of the void just next to the DLA. The top panel shows the synthetic spectrum without noise in red and the same spectrum with the Gaussian noise added. The noise level in the data is low enough to avoid noise spikes breaking the dark gap artificially.}
\label{fig:map}
\end{figure}

\subsection{So what gives?}

In order to understand better how the longest gaps form in the simulations, I show in Figure \ref{fig:map} a synthetic line of sight containing a gap of length $192h^{-1}\dim{cMpc}$. The line of sight passes mostly along the filaments (red colors), snaking in between large voids (blue colors), but also crosses a void next to a Damped Lyman-$\alpha$ system (DLA). If the synthetic spectrum was computed with using Doppler profiles for all pixels, the large gap would be split into two smaller gaps by a transmission spike through that void. It is, therefore, tempting to hypothesize that DLA play a role in "protecting" the long gaps from being terminated by nearby voids.

However, using Doppler instead of Voigt profiles for all synthetic spectra reduces the number of gaps in excess of $100h^{-1}\dim{cMpc}$ by only about 10\% (this is shown below in Fig.\ \ref{fig:mod}). Thus, having a DLA along the line of sight is not a necessary condition for having a large gap. 

If the high density regions harboring DLAs are not affecting the gap distribution significantly, then the only way to modify observables such as the gap distribution or the distributions of mean opacities is by modifying the transmission spikes themselves. To be specific, let's consider what needs to be done to bring the distribution of mean opacities in the DC=-1 run in agreement with observations. \citet{Zhu2021} rescaled all their simulations to match the observed mean opacity. In order to check the effect of such rescaling, I multiply all optical depths in the fully resolved synthetic spectra in the DC=-1 run uniformly by a factor of $f_\tau$ (which is equivalent to, for example, increasing the phototionization rate the factor of $1/f_\tau$). Varying $f_\tau$ between values of 1 and 0.75 smoothly interpolates between the DC=-1 and DC=0 runs in both the gap distribution and the distribution of mean opacities. In other words, synthetic spectra in the DC=-1 run with $f_\tau=0.75$ are very close to the synthetic spectra in the DC=0 run. That suggests that in CROC simulations the dark gap statistic is strongly correlated with the distribution of mean opacities and does not contain much additional information.

\begin{figure}[t]
\includegraphics[width=\hsize]{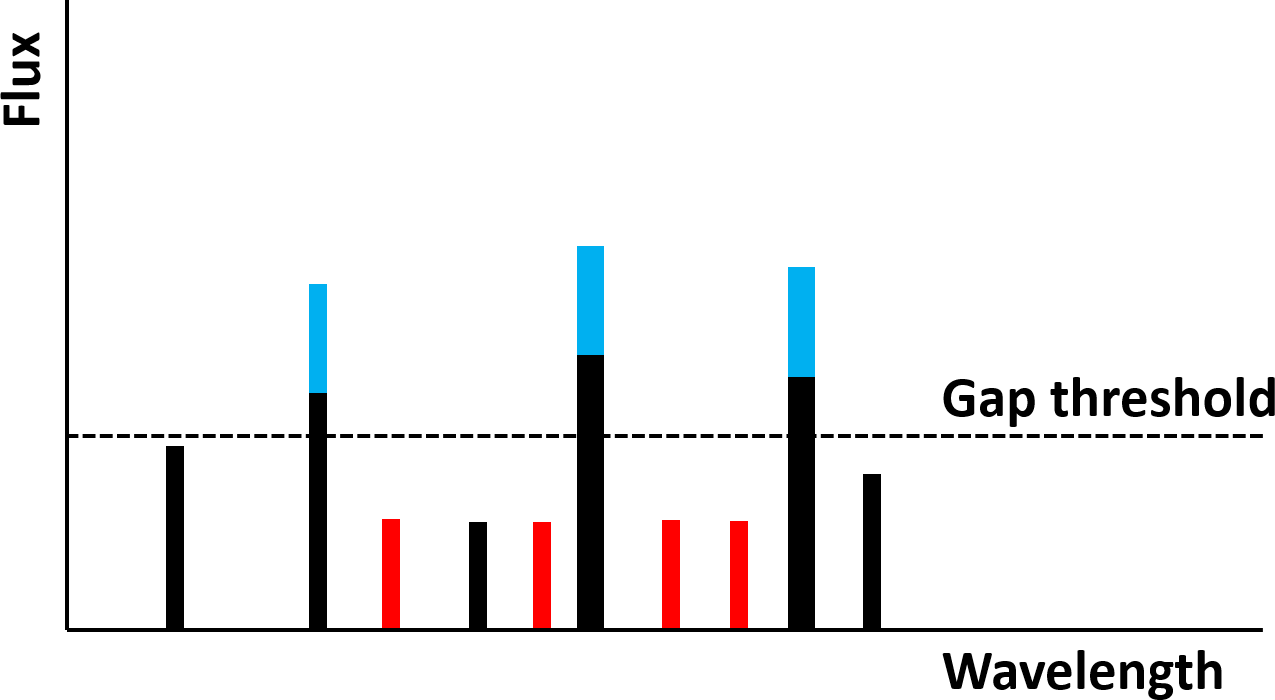}
\caption{Sketch of the possible "fixes" for reconciling the DC=-1 run with the observed distribution of mean opacities. Black bars sketch a piece of the spectrum in the original DC=-1 sim. Red lines show a "fix A", with more transmitted flux at high gas densities, below the gap detection threshold (so that the gap distribution is unaffected, but the mean opacity is). Blue lines are for deeper, more transparent voids, with more transmitted flux in already existing spikes that also does not affect the gap distribution.}
\label{fig:sk}
\end{figure}

In order to break this correlation, regions of different densities need to be rescaled by different amounts. This is sketched in Figure \ref{fig:sk}. In order to avoid breaking the agreement of the DC=-1 run with the observed gap distribution, either small transmission spikes below the gap detection threshold can be added (I call this "fix A") or the voids can be made deeper, with transmission spikes that cross the gap detection threshold being higher, thus decreasing the mean opacity and but again affecting the gap distribution ("fix B"). 

\begin{figure*}[t]
\includegraphics[width=0.5\hsize]{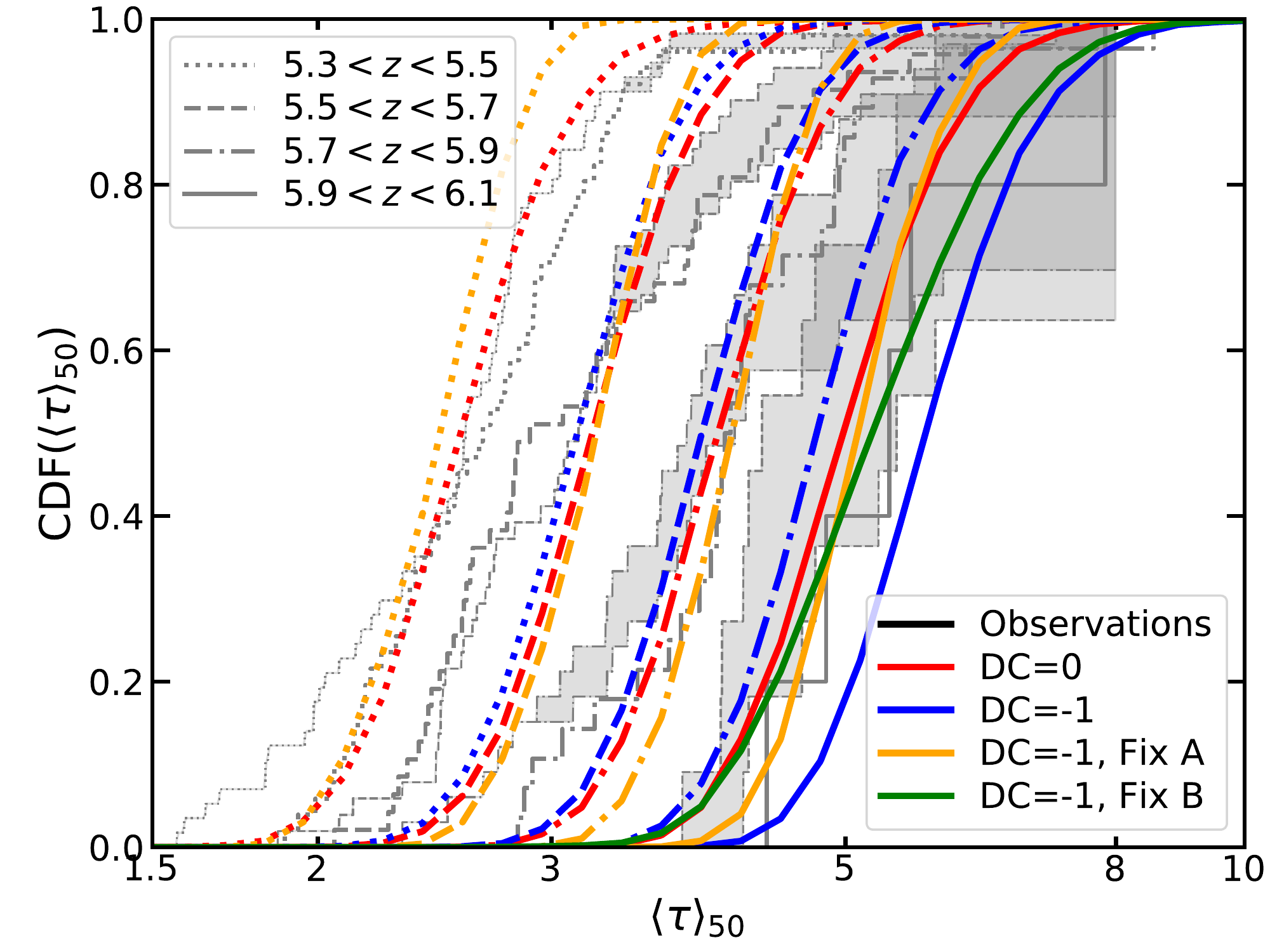}%
\includegraphics[width=0.5\hsize]{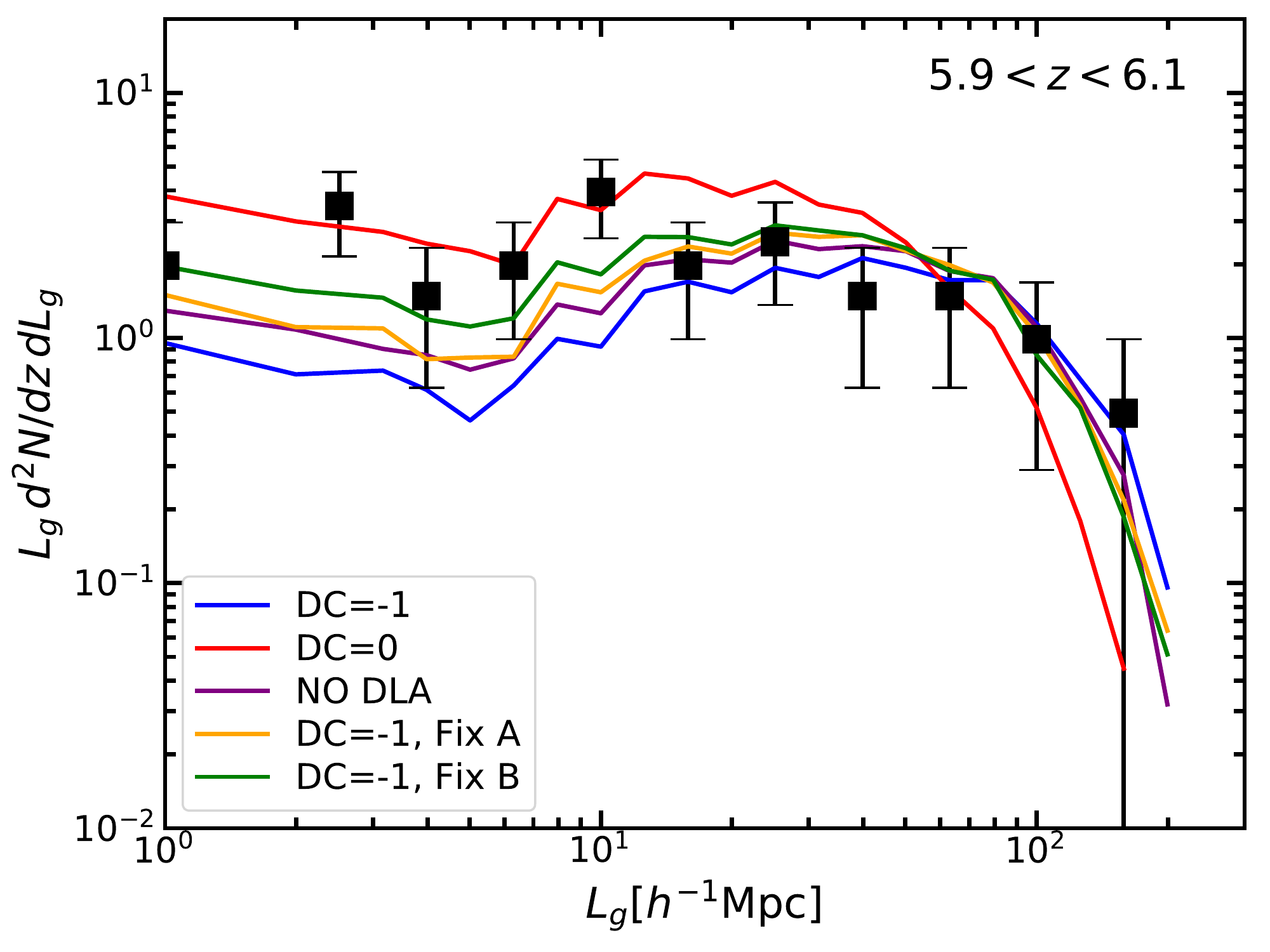}
\caption{Distributions of mean opacities in $50h^{-1}\dim{cMpc}$ skewers (left) and probability distribution functions for gaps of different lengths (right) for  per unit redshift for gaps whose central redshifts fall in the redshift bin $[5.9,6.1]$. Blue and red lines show the DC=-1 and DC=0 CROC simulations respectively as in the previous figures. The purple line (show in the right panel only) is for the DC=-1 run with Voigt profiles replaced by Doppler profiles in synthetic spectra. Orange and green lines are for the two artificial "fixes" for the DC=-1 run described in the text.}
\label{fig:mod}
\end{figure*}

Specific modifications I apply are decreasing the HI fraction by a factor of (a) 0.15 for all gas with baryon density $\rho>0.5\bar\rho$ for "fix A" and (b) 0.3 for all gas with baryon density $\rho<0.17\bar\rho$ for "fix B" (such "fixes" are obviously not physical and only serve as the demonstration of the possibility to modify the neutral hydrogen density distribution in the simulations to match both observational constraints simultaneously). Distributions of gaps and mean opacities for these two "fixes" are shown in Figure \ref{fig:mod} together with the original DC=0 and DC=-1 runs. There are issues with both "fixes", however.

"Fix A" makes the higher density regions more transparent, and while shifting the full CDF of mean opacities towards the observations, it makes the distribution \emph{narrower}, while the observed distributions are actually a bit (albeit insignificantly) wider than the simulated ones. It may still be consistent with the data, but checking that would require a careful analysis, including understanding uncertainties on the observed distributions.

"Fix B" makes the CDF of mean opacities wider, and at face value may seem like a good approach. However, it implies that the densities of voids are significantly underestimated in the simulations. While it is not inconceivable in principle, both CROC simulations and simulations of \citet{Zhu2021} exhibit good numerical convergence with respect to the spatial resolution or the velocity width of a pixel in the synthetic spectra, so any mis-prediction of densities in voids would imply false numerical convergence in existing simulations.

\section{Conclusions}

The main defect of CROC simulations is the incorrect abundance of super-$L_*$ galaxies \citep{Zhu2020croc}. Simulations do not make enough of them, and to compensate for the under-abundance of sources of ionizing radiation, the ionizing efficiencies of smaller galaxies are overestimated. Since the massive galaxies appear at lower redshifts, the overall effect of that defect is to make reionization proceed too fast and end too early in the simulations. Even with artificially early reionization the DC=-1 CROC runs provides a decent match to the observed distributions of long dark gaps (although fails to match gaps shorter than $10h^{-1}\dim{cMpc}$), implying that the distribution of long gaps by itself does not constrain the timing of reionization.

If the hypothesis proposed in this paper - that the distribution of gaps is primarily controlled by the ionization level in voids (from which high enough transmission spikes that terminate long gaps originate) - is indeed correct, then the distribution of gaps is expected to correlate strongly with the fraction of quasar spectra above the gap detection threshold (in the absence of additional instrumental broadening that can increase this fraction artificially). This is good news, since this fraction can easily be measured, while the abundance of voids has been proposed as a good probe of cosmic non-Gaussianity \citep{voidng1,voidng2,voidng3,voidng4,voidng5}. Hence, further exploration of the physical origin of long dark gaps may eventually develop into a novel probe of cosmology.

Each of the two CROC runs (DC=0 and DC=-1) match either the distribution of dark gaps or the distribution of mean opacities, but not both. Moreover, the DC=-1 run can be rescaled to match the distribution of mean opacities (either by increasing the photoionization rate by 33\% or the temperature by 50\%), in which case it becomes nearly indistinguishable from the DC=0 run and fails the $F_{30}$ diagnostic just as badly. One can "fix" the CROC simulations to match the both observational diagnostics simultaneously in two different ways, by either making higher densities more transparent or making voids deeper, but both "fixes" have their own difficulties and are not self-consistent. In short, it is currently not clear how to modify the physical model used by CROC simulations to achieve a simultaneous fit to both the dark gap distribution and the the distribution of mean opacities.

\acknowledgments
I am grateful to Andrey Kravtsov, Sergei Shandarin, Christine Simpson, and Huanqing Chen for multiple helpful suggestions and comments and to Yongda Zhu and George Becker for providing access to their data and for constructive and productive discussion.

This manuscript has been authored by Fermi Research Alliance, LLC under Contract No. DE-AC02-07CH11359 with the U.S. Department of Energy, Office of Science, Office of High Energy Physics. CROC project relied on resources of the Argonne Leadership Computing Facility, which is a DOE Office of Science User Facility supported under Contract DE-AC02-06CH11357. An award of computer time was provided by the Innovative and Novel Computational Impact on Theory and Experiment (INCITE) program. CROC project is also part of the Blue Waters sustained-petascale computing project, which is supported by the National Science Foundation (awards OCI-0725070 and ACI-1238993) and the state of Illinois. Blue Waters is a joint effort of the University of Illinois at Urbana-Champaign and its National Center for Supercomputing Applications. 

\appendix

\begin{figure*}[t]
\includegraphics[width=0.5\hsize]{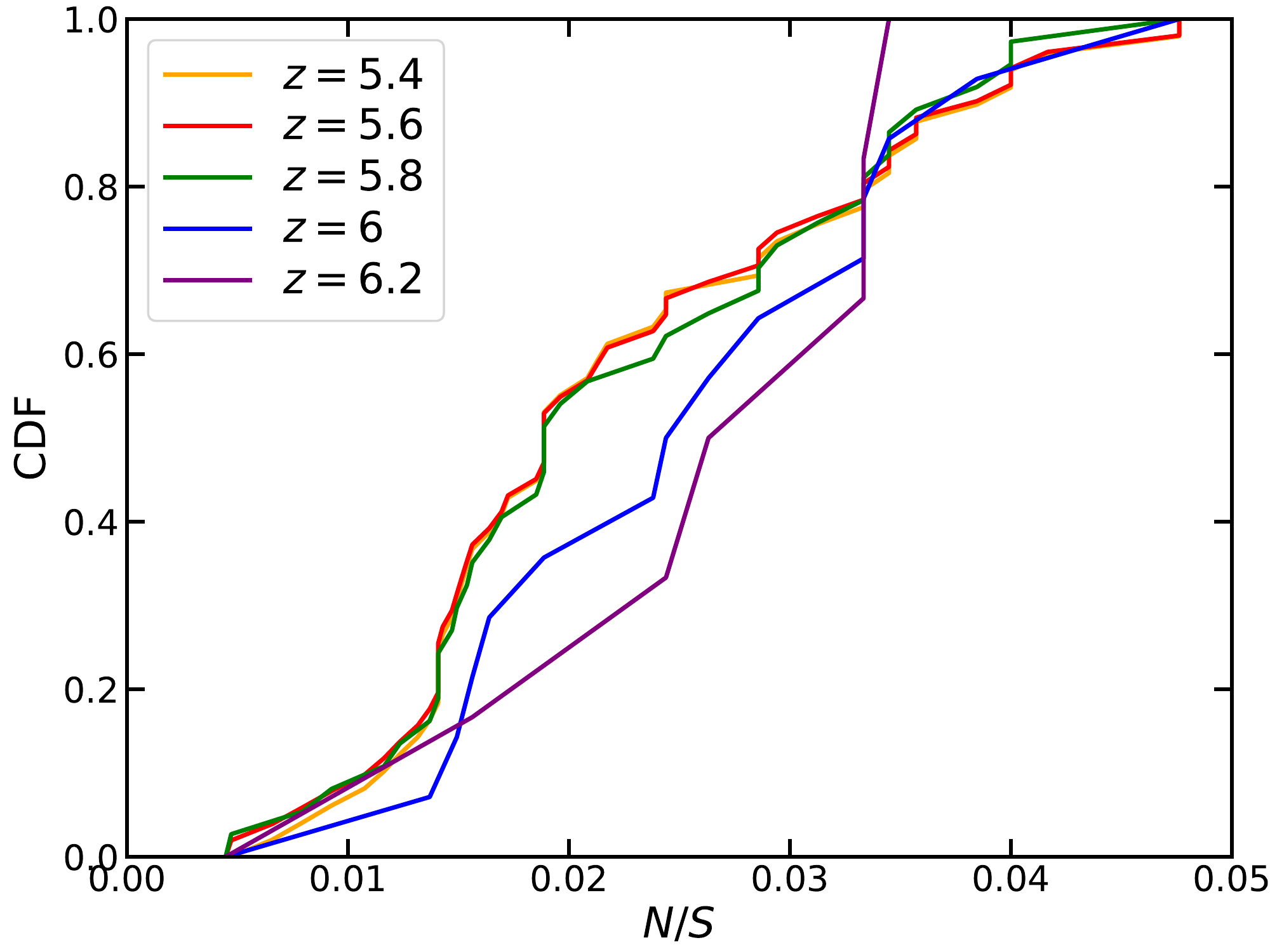}%
\includegraphics[width=0.5\hsize]{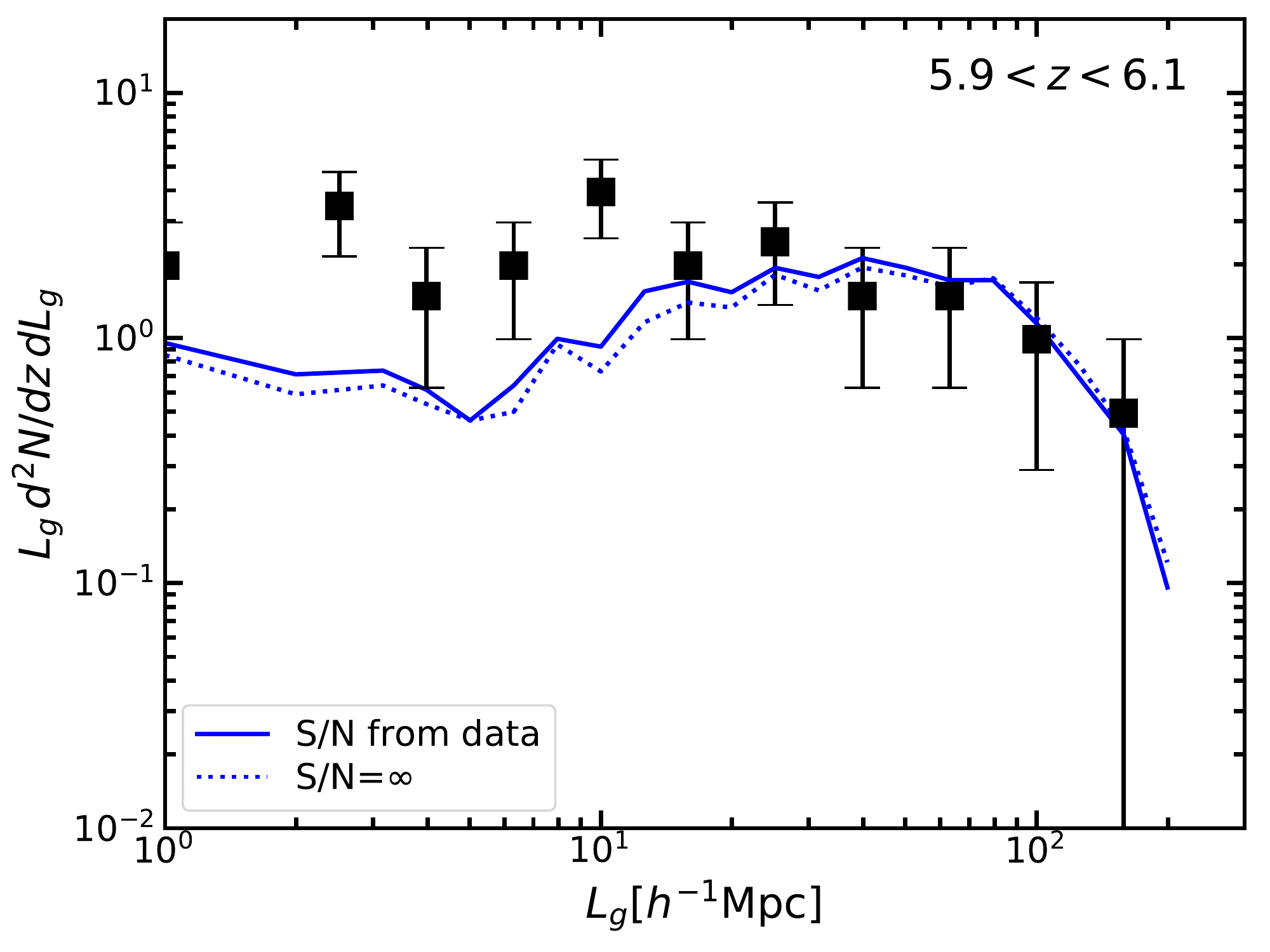}
\caption{Left: cumulative distributions for observational noise-to-signal ratios for the \citet{Zhu2021} sample at different redshifts. These distributions need to sampled properly in order to compare simulation results with the data. Right: dark gap distributions for the DC=-1 run with and without Gaussian noise at $z=6$. The noise level in the observational data is low enough not to affect the dark gap distribution appreciably.}
\label{fig:s2n}
\end{figure*}

The observational data include a range of quasar spectra with different signal-to-noise values, and at different redshifts a different subset of quasars contributes to the overall distribution of dark gaps. The left panel of Fig.\ \ref{fig:s2n} shows the cumulative distribution of signal-to-noise ratios of quasar spectra that intersect specific redshift values. These values are quoted at $30\,\dim{km/s}$ resolution ($R=10^4$), while the dark gaps are selected from spectra rebinned to $1h^{-1}\dim{cMpc}$ ($R=2000$). In forward modeling the simulated lines of sight I assume that noise in neighboring pixels is uncorrelated, so it is sufficiently accurate just to rescale values from the left panel of Fig.\ \ref{fig:s2n} by a factor $\sqrt{0.2}$. With such rescaling, the distributions of noise values at each redshift are sampled randomly so that the noise added to the synthetic spectra is consistent with the distribution of noise in the observational data. 

One may think that it matters how these distributions are sampled. For example, if a synthetic line of sight containing a particularly large gap is assigned high enough noise, a noise spike may break the large gap into two smaller ones. I have tested that and found that it is in fact not necessary to bootstrap the synthetic spectra, the 1000 lines of sight are enough to  sample the joint distribution of dark gaps and observational signal-to-noise ratios.

Dark gap distributions for synthetic lines of sight with and without Gaussian noise are shown in the right panel of Fig.\ \ref{fig:s2n}. The noise level in the observational data is low enough not to affect the dark gap distribution appreciably.

\bibliographystyle{aasjournal}
\bibliography{main}

\begin{thebibliography}{}
\expandafter\ifx\csname natexlab\endcsname\relax\def\natexlab#1{#1}\fi
\providecommand{\url}[1]{\href{#1}{#1}}

\bibitem[{{Achitouv} {et~al.}(2015){Achitouv}, {Neyrinck}, \&
  {Paranjape}}]{delv2}
{Achitouv}, I., {Neyrinck}, M., \& {Paranjape}, A. 2015, \mnras, 451, 3964

\bibitem[{{Baghram} {et~al.}(2013){Baghram}, {Namjoo}, \&
  {Firouzjahi}}]{voidng3}
{Baghram}, S., {Namjoo}, M.~H., \& {Firouzjahi}, H. 2013, \jcap, 2013, 048

\bibitem[{{Bardeen} {et~al.}(1986){Bardeen}, {Bond}, {Kaiser}, \&
  {Szalay}}]{BBKS}
{Bardeen}, J.~M., {Bond}, J.~R., {Kaiser}, N., \& {Szalay}, A.~S. 1986, \apj,
  304, 15

\bibitem[{Becker {et~al.}(2015)Becker, Bolton, Madau, Pettini, Ryan-Weber, \&
  Venemans}]{Becker2015}
Becker, G.~D., Bolton, J.~S., Madau, P., {et~al.} 2015, Monthly Notices of the
  Royal Astronomical Society, 447, 3402.
\newblock \url{https://doi.org/10.1093/mnras/stu2646}

\bibitem[{{Berg} {et~al.}(2019){Berg}, {Ellison}, {S{\'a}nchez-Ram{\'\i}rez},
  {L{\'o}pez}, {D'Odorico}, {Becker}, {Christensen}, {Cupani}, {Denney}, \&
  {Worseck}}]{Berg2019}
{Berg}, T. A.~M., {Ellison}, S.~L., {S{\'a}nchez-Ram{\'\i}rez}, R., {et~al.}
  2019, \mnras, 488, 4356

\bibitem[{Bosman {et~al.}(2018)Bosman, Fan, Jiang, Reed, Matsuoka, Becker, \&
  Haehnelt}]{Bosman2018}
Bosman, S. E.~I., Fan, X., Jiang, L., {et~al.} 2018, Monthly Notices of the
  Royal Astronomical Society, 479, 1055.
\newblock \url{https://doi.org/10.1093/mnras/sty1344}

\bibitem[{Chan {et~al.}(2014)Chan, Hamaus, \& Desjacques}]{Chan2014}
Chan, K.~C., Hamaus, N., \& Desjacques, V. 2014, Phys. Rev. D, 90, 103521.
\newblock \url{https://link.aps.org/doi/10.1103/PhysRevD.90.103521}

\bibitem[{{Chan} {et~al.}(2014){Chan}, {Hamaus}, \& {Desjacques}}]{delv1}
{Chan}, K.~C., {Hamaus}, N., \& {Desjacques}, V. 2014, \prd, 90, 103521

\bibitem[{{Dall'Aglio} \& {Gnedin}(2010)}]{DallAglio2010}
{Dall'Aglio}, A., \& {Gnedin}, N.~Y. 2010, \apj, 722, 699

\bibitem[{{D'Aloisio} {et~al.}(2013){D'Aloisio}, {Zhang}, {Shapiro}, \&
  {Mao}}]{voidng4}
{D'Aloisio}, A., {Zhang}, J., {Shapiro}, P.~R., \& {Mao}, Y. 2013, \mnras, 433,
  2900

\bibitem[{{D'Amico} {et~al.}(2011){D'Amico}, {Musso}, {Nore{\~n}a}, \&
  {Paranjape}}]{voidng1}
{D'Amico}, G., {Musso}, M., {Nore{\~n}a}, J., \& {Paranjape}, A. 2011, \prd,
  83, 023521

\bibitem[{Eilers {et~al.}(2018)Eilers, Davies, \& Hennawi}]{Eilers2018}
Eilers, A.-C., Davies, F.~B., \& Hennawi, J.~F. 2018, The Astrophysical
  Journal, 864, 53.
\newblock \url{https://doi.org/10.3847/1538-4357/aad4fd}

\bibitem[{{Fan} {et~al.}(2006){Fan}, {Strauss}, {Becker}, {White}, {Gunn},
  {Knapp}, {Richards}, {Schneider}, {Brinkmann}, \& {Fukugita}}]{Fan2006}
{Fan}, X., {Strauss}, M.~A., {Becker}, R.~H., {et~al.} 2006, \aj, 132, 117

\bibitem[{{Garaldi} {et~al.}(2021){Garaldi}, {Kannan}, {Smith}, {Springel},
  {Pakmor}, {Vogelsberger}, \& {Hernquist}}]{Garaldi2021}
{Garaldi}, E., {Kannan}, R., {Smith}, A., {et~al.} 2021, arXiv e-prints,
  arXiv:2110.01628

\bibitem[{{Gnedin}(2000)}]{Gnedin2000}
{Gnedin}, N.~Y. 2000, \apj, 535, 530

\bibitem[{{Gnedin}(2004)}]{Gnedin2004}
---. 2004, \apj, 610, 9

\bibitem[{{Gnedin}(2014)}]{gnedin14}
---. 2014, \apj, 793, 29

\bibitem[{{Gnedin} {et~al.}(2017){Gnedin}, {Becker}, \& {Fan}}]{gnedin_etal17}
{Gnedin}, N.~Y., {Becker}, G.~D., \& {Fan}, X. 2017, \apj, 841, 26

\bibitem[{{Gnedin} {et~al.}(2011){Gnedin}, {Kravtsov}, \& {Rudd}}]{Gnedin2011}
{Gnedin}, N.~Y., {Kravtsov}, A.~V., \& {Rudd}, D.~H. 2011, \apjs, 194, 46

\bibitem[{Hamaus {et~al.}(2014)Hamaus, Sutter, \& Wandelt}]{Hamaus2014}
Hamaus, N., Sutter, P., \& Wandelt, B.~D. 2014, Proceedings of the
  International Astronomical Union, 11, 538

\bibitem[{{Hamaus} {et~al.}(2016){Hamaus}, {Sutter}, \& {Wandelt}}]{Hamaus2016}
{Hamaus}, N., {Sutter}, P.~M., \& {Wandelt}, B.~D. 2016, in The Zeldovich
  Universe: Genesis and Growth of the Cosmic Web, ed. R.~{van de Weygaert},
  S.~{Shandarin}, E.~{Saar}, \& J.~{Einasto}, Vol. 308, 538--541

\bibitem[{{Hamaus} {et~al.}(2014){Hamaus}, {Wandelt}, {Sutter}, {Lavaux}, \&
  {Warren}}]{Hamaus2014b}
{Hamaus}, N., {Wandelt}, B.~D., {Sutter}, P.~M., {Lavaux}, G., \& {Warren},
  M.~S. 2014, \prl, 112, 041304

\bibitem[{{Jennings} {et~al.}(2013){Jennings}, {Li}, \& {Hu}}]{Jennings2013}
{Jennings}, E., {Li}, Y., \& {Hu}, W. 2013, \mnras, 434, 2167

\bibitem[{{Kannan} {et~al.}(2021){Kannan}, {Garaldi}, {Smith}, {Pakmor},
  {Springel}, {Vogelsberger}, \& {Hernquist}}]{Kannan2021}
{Kannan}, R., {Garaldi}, E., {Smith}, A., {et~al.} 2021, \mnras,
  arXiv:2110.00584

\bibitem[{{Pen}(1997)}]{Pen1997}
{Pen}, U.-L. 1997, \apjl, 490, L127

\bibitem[{{S{\'a}nchez-Ram{\'\i}rez} {et~al.}(2016){S{\'a}nchez-Ram{\'\i}rez},
  {Ellison}, {Prochaska}, {Berg}, {L{\'o}pez}, {D'Odorico}, {Becker},
  {Christensen}, {Cupani}, {Denney}, {P{\^a}ris}, {Worseck}, \&
  {Gorosabel}}]{Sanchez-Ram2016}
{S{\'a}nchez-Ram{\'\i}rez}, R., {Ellison}, S.~L., {Prochaska}, J.~X., {et~al.}
  2016, \mnras, 456, 4488

\bibitem[{{Sekiguchi} \& {Yokoyama}(2012)}]{voidng2}
{Sekiguchi}, T., \& {Yokoyama}, S. 2012, arXiv e-prints, arXiv:1204.2726

\bibitem[{{Sheth} \& {van de Weygaert}(2004)}]{Sheth2004}
{Sheth}, R.~K., \& {van de Weygaert}, R. 2004, \mnras, 350, 517

\bibitem[{{Sirko}(2005)}]{Sirko2005}
{Sirko}, E. 2005, \apj, 634, 728

\bibitem[{{Uhlemann} {et~al.}(2018){Uhlemann}, {Pajer}, {Pichon}, {Nishimichi},
  {Codis}, \& {Bernardeau}}]{voidng5}
{Uhlemann}, C., {Pajer}, E., {Pichon}, C., {et~al.} 2018, \mnras, 474, 2853

\bibitem[{Yang {et~al.}(2020)Yang, Wang, Fan, Hennawi, Davies, Yue, Eilers,
  Farina, Wu, Bian, Pacucci, \& Lee}]{Yang2020}
Yang, J., Wang, F., Fan, X., {et~al.} 2020, The Astrophysical Journal, 904, 26.
\newblock \url{https://doi.org/10.3847/1538-4357/abbc1b}

\bibitem[{{Zhu} {et~al.}(2020){Zhu}, {Avestruz}, \& {Gnedin}}]{Zhu2020croc}
{Zhu}, H., {Avestruz}, C., \& {Gnedin}, N.~Y. 2020, \apj, 899, 137

\bibitem[{{Zhu} {et~al.}(2021){Zhu}, {Becker}, {Bosman}, {Keating},
  {Christenson}, {Ba{\~n}ados}, {Bian}, {Davies}, {D'Odorico}, {Eilers}, {Fan},
  {Haehnelt}, {Kulkarni}, {Pallottini}, {Qin}, {Wang}, \& {Yang}}]{Zhu2021}
{Zhu}, Y., {Becker}, G.~D., {Bosman}, S. E.~I., {et~al.} 2021, \apj, 923, 223

\end{thebibliography}

\end{document}